# Optical photon detection in Al Superconducting Tunnel Junctions

G. Brammertz[a], A. Peacock[a], P. Verhoeve[a], D. Martin[a], R. Venn[b]

[a]*Science Payload and Advanced Concepts Office, ESA/ESTEC, Postbus 299, 2200AG Noordwijk, The Netherlands.*

[b]*Cambridge MicroFab Ltd., Trollheim Cranes Lane, Kingston, Cambridge CB3 7NJ, UK.*

**Abstract**

We report on the successful fabrication of low leakage aluminium superconducting tunnel junctions with very homogeneous and transparent insulating barriers. The junctions were tested in an adiabatic demagnetisation refrigerator with a base temperature of 35 mK. The normal resistance of the junctions is equal to ~7 $\mu\Omega$ cm$^2$ with leakage currents in the bias voltage domain as low as 100 fA/$\mu$m$^2$. Optical single photon counting experiments show a very high responsivity with charge amplification factors in excess of 100. The total resolving power $\lambda/\Delta\lambda$ (including electronic noise) for 500 nm photons is equal to 13 compared to a theoretical tunnel limited value of 34. The current devices are found to be limited spectroscopically by spatial inhomogeneities in the detectors response

Keywords: Superconducting Tunnel Junction; Al; optical photon detectors

*PACS*: 85.25.Oj; 85.25.Am; 85.25.-j

## 1. Introduction

The number of quasiparticles created in a Superconducting Tunnel Junction (STJ) is inversely proportional to the gap energy of the absorbing electrode, leading to an improved FWHM energy resolution for materials with a lower energy gap:

$$\Delta E = 2.355\sqrt{1.7\Delta_g (F+G+H)E} , \qquad (1)$$

where $\Delta_g$ is the energy gap of the absorbing electrode, E is the photon energy, F is the Fano factor [1,2], G is the statistical broadening due to multiple tunnelling of the quasiparticles [3,4,5,6] and H is the factor due to cancellation tunnel events [7,8], which become important for junctions with a low energy gap. In view of the better predicted energy resolution compared to Ta-based STJs ($\Delta_g$=700µeV) we have successfully fabricated high-quality Al-based ($\Delta_g$=180µeV) junctions and tested their abilities as NIR to soft-UV photon detectors.

## 2. Junction fabrication and experimental set-up

All junctions were fabricated by Cambridge MicroFab Ltd. The 100 nm thick base electrode is DC-sputtered onto a sapphire substrate at liquid nitrogen temperature. Without breaking vacuum a thin (~1nm) Al oxide layer is grown on top of the film followed by the deposition of the 50 nm top electrode. Both base and top film are polycrystalline with a residual resistance ratio of ~10. The tri-layer is patterned using UV-lithography and covered by 350 nm of reactively sputtered Si oxide. The contacts to the top film and the 3x3 µm$^2$ plug in the base lead are made out of Nb in order to prevent diffusion of the quasiparticles out of the junction area. Fig. 1 shows a schematic cross-section of the final junction. The devices have a square geometry with lateral device sizes ranging from 10 to 70 µm.

The junctions are operated in a two-stage adiabatic demagnetisation refrigerator with a base temperature of about 35 mK. The junctions are optically coupled via an optical fibre to a Xenon lamp in combination with a double grating monochromator, able to produce photons having a wavelength varying from 250 to 1000 nm. The illumination of the junctions is made through the back of the chip via the sapphire substrate.

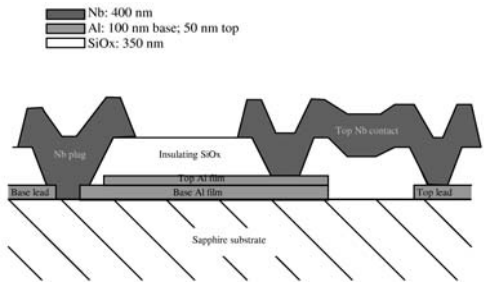

Fig. 1. Schematic cross-section of an Al-based STJ showing from left to right: the Al base lead, the Nb base lead plug, the junction tri-layer, the Nb top contact and the Al top lead.

For optical photon detection experiments the junctions are read out through a room-temperature charge-sensitive preamplifier followed by a shaping stage. The charge sensitive preamplifier integrates the current pulse created by the photo-absorption process in the junction. The output signal of the preamplifier is then fed into the shaping stage, consisting of two bipolar semi-gaussian shaping filters with fixed centre frequencies of respectively 5 and 33 kHz. In parallel to the photo-signal, an electronic pulser can be fed into the preamplifier, in order to measure the electronic noise

## 3. IV-characteristics

The insulating barrier between the electrodes is uniform and of high quality. Fig. 2 shows an IV-curve of the sub-gap region of a 30 μm side length junction taken at 35 mK with an applied parallel magnetic field of 50 Gauss. Leakage currents are as low as 100 fA per μm$^2$ of junction area and the dynamical resistivity $\rho_d$ in the bias domain is equal to ~4.5 Ωcm$^2$. On the other hand the normal resistivity $\rho_{nn}$ of the barrier is equal to ~7 μΩcm$^2$, giving a quality factor $Q=\rho_d/\rho_{nn}=6.4\ 10^5$.

## 4. Optical photon detection

All the junctions were illuminated with NIR to soft UV photons (λ=250-1000nm). The applied bias voltage is typically 100μV. Fig. 3 shows a collage of 5 different spectra acquired with the 30x30 μm$^2$ junction. The charge output and decay time of the Al junctions show a strong dependence on device size (Fig. 4), indicating that the losses are predominantly at the edges of the junction and/or into the leads and not in the bulk of the Al. Considering that approximately 3300 quasiparticles are created per eV of incoming photon energy in an Al electrode[9], we can conclude that the amplification due to multiple tunnelling of the quasiparticles varies between 7 and 105.

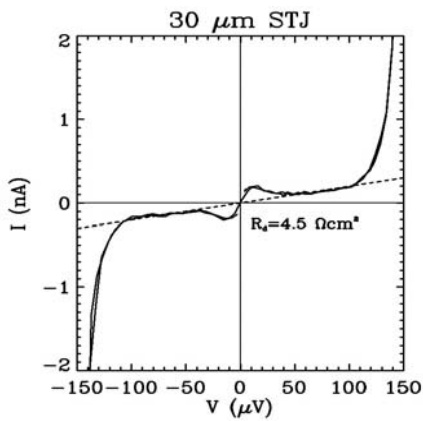

Fig. 2. IV-curve of the sub-gap region of a 30x30 μm$^2$ Al STJ.

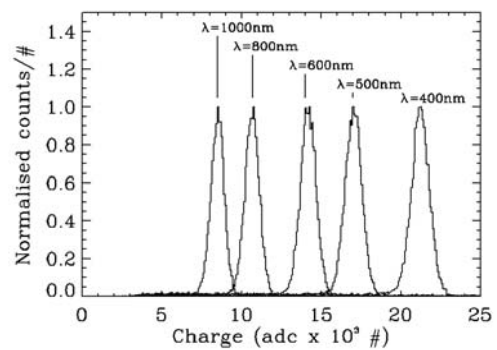

Fig. 3. Collage of five different spectra acquired with the 30 μm

The energy resolution of the junctions has not yet reached the expected tunnel-limited value for a pure Al junction. Fig. 5 shows the measured photo-peak width, the electronic pulser width, and the deduced intrinsic peak width as a function of incoming photon energy for the 30 μm side length device. The solid curve is the expected tunnel-limited resolution of Al STJs given by (1), where F is chosen equal to 0.22 and the tunnel term G equal to 1, which is the value for multiple tunnelling junctions with infinite integration time of the current pulse. The cancellation term H is in our case equal to 0.22 ($\bar{n} = 81$ and $\gamma = 0.385$). The intrinsic peak width we observe lies well above this curve. The dashed-dotted line is a least-squares fit to the intrinsic resolution using a function of the form:

$$\Delta E = 2.355\sqrt{1.7\Delta_g (F+G+H)E + \alpha E^2} \quad , \quad (2)$$

where $\alpha$ is a free fitting parameter representing disuniformities in the response across the junction. The deduced value for $\alpha$ is $\sim 4\cdot 10^{-4}$. We recall here that the best Ta based junctions show values for $\alpha$ equal to $2\cdot 10^{-6}$, revealing that there is still considerable room for improvement[10]. The resolution broadening in the Al junctions is likely to be related to the losses into the leads and at the edges of the detectors.

## 5. Conclusions

We have fabricated low-leakage Al STJs and operated them as optical photon detectors at temperatures below 100mK. The charge amplification due to multiple tunnelling varies between 7 and 105 depending on device size. The energy resolution has not yet reached the expected tunnel limited resolution for an Al STJ, because of spatial inhomogeneities in the response of the detector. Nevertheless, a measured resolving power of 13 for 500 nm photons has already been achieved.

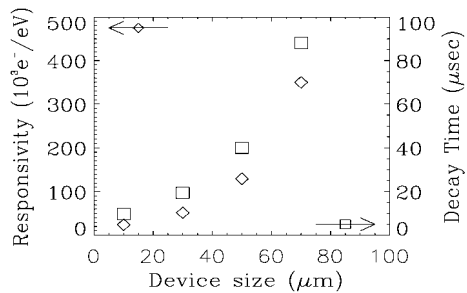

Fig. 4. Responsivity and decay time as a function of device size.

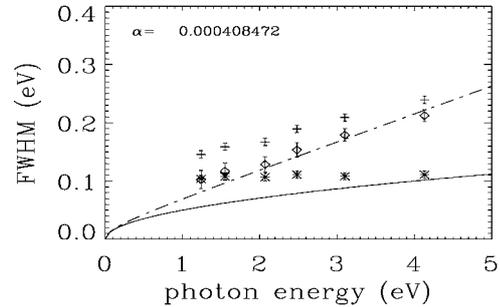

Fig. 5. Measured photo peak width (crosses), electronic pulser width (stars) and deduced intrinsic peak width (diamonds) for the 30 μm side length junction. The solid line is the expected tunnel limited energy resolution of Al STJs. The dashed-dotted line is a least squares fit to the intrinsic peak width using a linear broadening contribution characterized by α.